\documentstyle[aps,preprint]{revtex}
\begin{document}
\draft
\title{ Conductance Peak Distributions in Quantum Dots and
  the Crossover between  Orthogonal and Unitary Symmetries}

\author{ Y. Alhassid$^{1,2}$, J. N. Hormuzdiar$^1$ and N.D. Whelan$^3$}

\address{$^1$ Center for Theoretical Physics, Sloane Physics Laboratory,
	 Yale University, New Haven, Connecticut 06520\\
  $^2$ Insitute for Theoretical Physics, University of California, 
  Santa Barbara, California 93106\\
	 $^3$ Division de Physique Th\'{e}orique\cite{cnrs},
Institut de Physique
          Nucl\'{e}aire, 91406  Orsay Cedax, France \\
          and CATS, Niels Bohr Institute, 17 Blegdamsvej Copenhagen, Denmark }

\date {\today}
\maketitle
\begin{abstract}
 Closed expressions are derived for the resonance widths and Coulomb
blockade conductance peak heights in quantum dots for the crossover 
regime between conserved and broken time-reversal symmetry.
 The results hold for leads with any number of possibly correlated and 
inequivalent channels. Our analytic
predictions are in good agreement with simulations of both random matrices
and a chaotic billiard with a magnetic flux line. 
\\
\end{abstract}

\pacs{PACS numbers: 73.40.Gk, 05.45+b, 73.20.Dx, 24.60.-k }
\narrowtext

  Recent advances in submicron technology have made it possible to 
fabricate ballistic quantum dots 
 in which the electron mean free path exceeds the system size.
For dots with irregular shapes, the  dynamics
 is largely chaotic \cite{Bo91} and its signatures 
 are observed in the transport properties of the device \cite{Me95}. 
 In  dots that are weakly coupled to external leads via tunnel
 barriers,  the resonances are  isolated, and
 at  low temperatures  the  conductance  is dominated by the
 resonance that is closest to  the Fermi energy  in the leads. 
Such closed microstructures thus offer a unique opportunity to  probe
 the chaotic signatures of individual wavefunctions.
Since the charging energy of the dot is large compared with the mean-level 
spacing, the conductance exhibits a series of almost equally spaced peaks
versus  the gate voltage  (Coulomb blockade oscillations).
 The peaks show order-of-magnitude fluctuations which 
are explained by a statistical theory \cite{JSA92,PEI93}.  The distributions
of the conductance peak heights were measured  for both conserved and 
broken time-reversal symmetry \cite{Chang96,Marcus96}, and were found to agree
 with theory. Using  the supersymmetry technique for disordered
 systems \cite{MPA95,Prigodin95} and random matrix theory (RMT) \cite{AL95},
 closed expressions for width and conductance
 peaks  distributions were derived for any number of  possibly  correlated 
 channels  for both orthogonal and unitary  symmetries.

 In this letter we derive exact  expressions for the universal width and
 conductance peak distributions in the crossover regime between 
conserved and broken
 time-reversal symmetry for any number of possibly correlated and/or 
inequivalent channels.
 An approximate expression for the distribution of the
 wavefunction intensity at a fixed spatial point was obtained in Ref. 
\cite{ZL91}, and an exact expression  was derived in Ref. \cite{FE94}
 using  supersymmetry. The  averaged intensity
 over the complete spectrum of the random matrix  was derived 
 in Ref.  \cite{Sommers94}.  

   At low temperatures  $\Gamma \ll kT \ll \Delta$, the conductance peak 
amplitude $G$ 
 is given by \cite{Be91} $G =\frac{e^2}{h}\, \frac{\pi \bar{\Gamma}}{4 kT} g$
where
\begin{eqnarray}\label{Peak}
   g=  {2 \over \bar{\Gamma}} \frac{\Gamma^l \Gamma^r}{\Gamma^l
 + \Gamma^r} \;.
\end{eqnarray}
Here $\Gamma^{l(r)}$  is the width
of a resonance to decay into  the
  left (right) lead and $\bar{\Gamma}$ is the total average width. 
 $g$ is dimensionless and  temperature-independent.
 In general we assume that the left (right) lead has  $\Lambda^{l(r)}$ open 
channels such that $\Gamma^{l(r)} = \sum_c |\gamma_{c}^{l(r)}|^2$, where
$\gamma_{c}^{l(r)}$ is the partial amplitude to decay into channel
$c$ on the left (right).  
 Using $R$-matrix theory\cite{JSA92} or resonance theory\cite{AA96}, 
 the partial  width amplitude can be expressed in terms of the resonance
 wavefunction $\Psi$  and the channel wavefunction  $\Phi_c$ 
at the lead-dot interface. When
  each lead is modelled in terms of several point-like contacts 
 ${\bf r}_c$ then $ \gamma_c \propto \Psi({\bf r}_c)$ \cite{MPA95} . 
 By expanding a resonance wavefunction 
 in a complete basis of solutions $\rho_\mu(\vec{r})$
 inside the dot  at a fixed energy 
($\Psi({\bf r})= \sum_\mu \psi_\mu \rho_\mu({\bf r})$), 
the partial width $\gamma_{c}$  can
 be expressed as a scalar product \cite{AA96} of the vectors that represent the
channel and the resonance function  $\gamma_c =
 \langle \mbox{\boldmath$\phi$}_c | \mbox{\boldmath$\psi$} \rangle
 \equiv  \sum_\mu \phi_{c \mu}^\ast \psi_\mu$.

 In the crossover regime of breaking time-reversal symmetry, 
the resonance wavefunctions  are assumed to
have statistical properties which are described by the corresponding
 eigenvectors $\mbox{\boldmath$\psi$}$ of a
 random matrix ensemble that interpolates between the GOE and GUE \cite{Bo91}

\begin{eqnarray}\label{TranEns}
  H=S+i\alpha A 
\;.
\end{eqnarray}
 $S$ and $A$ are, respectively, symmetric and antisymmetric real
 matrices of dimension $N$ 
which are uncorrelated
 and chosen from gaussian ensembles of variance $a^2$. 
 The proper transition parameter that describes the crossover from
 GOE to GUE
is given by the ratio of the rms of a typical symmetry-breaking matrix element
 to the mean-level spacing  $\Delta$ \cite{FKPT85}

\begin{eqnarray}\label{tranpar}
\lambda = {\alpha a \over \Delta} = {\alpha \sqrt{N} \over \pi} \;,
\end{eqnarray}
 where we have used the mean level 
density at the middle of the spectrum.
 The complete breaking of time-reversal symmetry occurs for 
$\lambda  \sim1$. Since  the  transition parameter 
depends on the level density, we shall study the ensemble's statistics
 only around the middle of the spectrum.  The spectral properties of the
 transition ensemble
(\ref{TranEns}) were  derived in a closed form  \cite{PM83}.  Similar spectral
correlators were also derived for a single electron in
 a disordered medium using the supersymmetry method\cite{AIE93}. 
  However, less is known
 about the eigenvector statistics and, unlike in the GOE and GUE limits,
 the eigenvalue and eigenvector distributions do not factorize.

An eigenvector component $\psi_\mu$ 
 is a complex number, 
$\psi_\mu = \psi_{\mu R} +  i \psi_{\mu I} $,
and can be viewed as a two-dimensional vector  in the complex
 plane. Since the eigenvector $\mbox{\boldmath$\psi$}$ is determined only up to
 a phase $e^{i \theta}$,  $\psi_\mu$ is determined only up to a rotation by 
an angle $\theta$.  This angle is uniquely determined by rotating to 
 a principal frame in which \cite{FKPT85,ZL91} 

\begin{eqnarray}\label{scalars}
\sum\limits_{\mu=1}^N \psi_{\mu R} \psi_{\mu I}  = 0 
\;\;;\;\;\;   \sum\limits_\mu 
\psi_{\mu I}^2  /  \sum\limits_\mu \psi_{\mu R}^2 \equiv r^2 
\end{eqnarray}
with $0\leq r \leq1$.
The ratio $r^2$ is invariant under a rotation and is thus independent of the
frame. The  right inset  of Fig. 1 shows the components
 of a typical eigenvector  in the complex plane. Its general  shape is
that of an  ellipsoid whose semi-axes define the principal axes.
 In the following,   $\psi_{\mu R}$ and $\psi_{\mu I}$ will
denote exclusively the components of $\mbox{\boldmath$\psi$}$
 in this principal frame.  

    Let us consider eigenvectors with a  fixed  value 
of $r$. Under an orthogonal  transformation (in the $N$-dimensional space)
 the real and imaginary parts of $\mbox{\boldmath$\psi$}$ do
 not mix  so that a principal frame
remains principal and $r^2$ is invariant.  Since the probability
 distribution of the 
ensemble (\ref{TranEns}) is also invariant under an orthogonal transformation,
 the conditional probability of finding an eigenvector
 $ \mbox{\boldmath$\psi$} $  given its $r$-value is 

\begin{eqnarray}\label{psiDist}
  P(\mbox{\boldmath$\psi$} | r) \propto 
    \delta  \left( \sum\limits_\mu  \psi_{\mu R}^2 -{1  \over1+r^2} \right)
 \delta \! \left( \sum\limits_\mu \psi_{\mu I} ^2 - { r^2 \over1+r^2} \right)
  \delta \! \left( \sum\limits_\mu \psi_{\mu R} \psi_{\mu I} \right) \;.
\end{eqnarray}
    The conditional distribution $P(X | r)$ of any quantity $X$ which is 
a function of the eigenvector
 $\mbox{\boldmath$\psi$}$  can be calculated from (\ref{psiDist}). 
  The actual distribution  $P_\lambda(X)$  at a given value $\lambda$ of
 the transition parameter  can then be calculated in terms of  
the distribution $P_\lambda(r)$ of  the quantity  $r$ (whose explicit form
 is given below in Eq. (\ref{rDist}))

\begin{eqnarray}\label{fullDist}
P_\lambda(X) = \int\limits_0^1 P_\lambda(r)  P(X | r) \equiv 
\langle P(X | r) \rangle
\;.
\end{eqnarray}

 The conditional  joint distribution of the partial  width  amplitudes
$\mbox{\boldmath$\gamma$} =(\gamma_1, \gamma_2,  \ldots,
\gamma_\Lambda)$ for $\Lambda$  real channels $\mbox{\boldmath$\phi$}_c$
 can be calculated exactly from (\ref{psiDist}) and $\gamma_c=\langle
\mbox{\boldmath$\phi$}_c | \mbox{\boldmath$\psi$} \rangle$.
   In the limit $N\rightarrow \infty$  we obtain a  Gaussian distribution
 in $\gamma_{cR}$ and $\gamma_{cI}$

\begin{eqnarray}\label{gammaDist}
  P(\mbox{\boldmath$\gamma$} |r ) \propto
\exp \left( {1+ r^2 \over 2} \gamma^T_R M^{-1}
\gamma_R +{ 1+ r^2 \over 2 r^2} \gamma^T_I M^{-1} \gamma_I \right) 
\;,
\end{eqnarray}
where   $\gamma_c =\gamma_{c R} + i \gamma_{c I} =
 \langle \mbox{\boldmath$\phi$}_c | \psi_R \rangle +
 i  \langle \mbox{\boldmath$\phi$}_c | \psi_I \rangle$
is a principal frame decomposition,
and $\mbox{\boldmath$\gamma$}^T$ denotes the transpose of 
the column vector $\mbox{\boldmath$\gamma$}$.
  The matrix  $M$ is the channel  correlation matrix  $M_{c c^\prime} =
 \overline{\gamma_c^\ast \gamma_{c^\prime}} =
 \langle \mbox{\boldmath$\phi_c$} |  \mbox{\boldmath$\phi$}_{c^\prime} \rangle$.
To calculate $M$  for  
 a chaotic system, we note that the relation 
  $\overline{\psi_\mu^\ast \psi_{\mu^\prime}}
 = N^{-1}\delta_{\mu \mu^\prime}$ is valid for the transition ensemble
 irrespective of the value of $\lambda$.
We can therefore repeat the derivation of Ref. \cite{AL95} to find 
 for $M$ an expression identical to the one obtained in either the
 orthogonal or unitary limits. For point-like contacts 
 $M_{cc^\prime} \propto J_0(k|{\bf r}_c - {\bf r}_c^\prime|)$, 
independently  of the crossover magnetic field. 

 As a special case  of (\ref{gammaDist}) we obtain the conditional
 distribution of a single component  $\psi_\mu$ of an eigenvector 
by choosing a single channel along the $\mu$-th axis.
 At a fixed $r$, $\psi_{\mu R}$ and $\psi_{\mu I}$ are independent
 Gaussian variables with
 $\overline{\psi_{\mu I}^2} / \overline{\psi_{\mu R}^2 }= r^2$.
The full distributions of the real and imaginary parts of 
the wavefunction amplitude
are obtained by a weighted average of the conditional distributions
according to Eq. (\ref{fullDist}) and are not statistically independent.
   Fig. \ref{fig1} shows these distributions for $\lambda=0.1$.  A Gaussian 
distribution (dashed line) is a good approximation for the real part of the 
wavefunction amplitude,
  but not for the imaginary part.  

  The distribution of the width
$\Gamma= | \gamma |^2 = \gamma_R^2 + \gamma_I^2$ for a one-channel lead
is found to be

\begin{eqnarray}\label{GammaDist}
 P_\lambda(\hat{\Gamma}) = \left\langle a_+ e^{-a_+^2 \hat{\Gamma}} 
I_0 \left( a_+a_- \hat{\Gamma} \right) \right\rangle
\;,
\end{eqnarray}
where  $\hat{\Gamma} = \Gamma / \bar{\Gamma}$, $a_\pm 
 \equiv (r^{-1} \pm r)/2$
and $I_0$ is the modified Bessel function of order zero.
An exact expression for the distribution of the wavefunction intensity
(at a fixed spatial point) was derived in Ref. \cite{FE94} through the 
supersymmetry method. 
By comparing  our result (\ref{GammaDist}) with the result in \cite{FE94},
  we find that the  $r$-distribution is given by

\begin{eqnarray}\label{rDist}
 P_\lambda(r)  = \pi^2 \lambda^2 (1/r^3 -r ) 
 e^{-{\pi^2 \over 2} \lambda^2 \left(r- 1/r \right)^2}
 \left\{ \phi_1(\lambda)  + [ ( r + 1/r )^2 /4
 - 1/2\pi^2  \lambda^2] [1- \phi_1(\lambda) ] \right\} \;,
\end{eqnarray}
where $\phi_1(\lambda)= 
 \int\limits_0^1 e^{-2 \pi^2 \lambda^2(1-t^2)} dt$.  RMT simulations confirm 
that Eq. (\ref{rDist}) is indeed the $r$-distribution (see left inset in
Fig. \ref{fig1}).   Eqs. (\ref{gammaDist}), (\ref{fullDist})  and (\ref{rDist})
provide a closed analytic expression
 for the joint partial amplitudes distribution for a lead with 
 any number of  possibly correlated and inequivalent  channels. 

  To derive a closed expression for the total width distribution for
 multi-channels leads, we note that 
the conditional distribution $P(\mbox{\boldmath$\gamma$} | r)$ in
 Eq. (\ref{gammaDist}) is identical
to a GOE distribution for $2\Lambda$ channels 
with partial amplitudes $\gamma_{cR}, \gamma_{cI}$ and a correlation matrix 
${\cal M}$ composed of four $\Lambda \times \Lambda$ blocks 
\begin{eqnarray}\label{DoubleM}
{\cal M} = \left ( \begin{array}{cc}
                 {1 \over 1+r^2} M  & 0\\
                    0  &  {r^2 \over 1+r^2} M
         \end{array} \right) \;.
\end{eqnarray}
 We can therefore use the known distributions from the GOE limit \cite{AL95}.  
 The $2\Lambda$ eigenvalues of ${\cal M} $ are given by
$ \{ \omega^2_j \} = \{ {1 \over 1+r^2} w_c^2, {r^2 \over 1+r^2} w_c^2 \}$ where
 $w_c^2$ are the $\Lambda$ eigenvalues of $M$. Sorting the  inverse
 eigenvalues of ${\cal M}$ in
ascending order $\omega_1^{-2}< \omega_2^{-2} < \ldots$, we have

\begin{eqnarray}
\label{generalGamma}
 P_\lambda(\Gamma) = \left\langle\frac{1}{\pi 2^{\Lambda}}
 \left(\prod_c  \frac{1}{\omega_c} \right)
 \sum_{m=1}^{\Lambda} \int_{1/2 \omega^2_{2m-1}}^{1/2 \omega^2_{2m}}
 d \tau\right. 
\left.  \frac{\mbox{e}^{-\Gamma \tau}}
 {\sqrt{\prod_{r=1}^{2m-1}   (\tau - \frac{1}{2 \omega^2_r})
	\prod_{s=2m}^{2\Lambda} (\frac{1}{2 \omega^2_{s}} - \tau)}}
\right\rangle
 \;.
\end{eqnarray}

  Similarly we can use the  closed expression for
 the  GOE conductance distribution \cite{AL95} to obtain 
 $P(g|r)$ analytically and then use (\ref{fullDist}) to find
 $P_\lambda(g) = \langle P(g|r) \rangle$. In this calculation
 of $P(g|r)$ (where Eq. (\ref{Peak}) is exploited),  we take
advantage of the statistical
 independence of the conditional distributions
of the total widths in the left and right leads, i.e. 
$P(\Gamma^l,\Gamma^r | r) =
P(\Gamma^l |r) P( \Gamma^r |r)$. We note however that the 
widths themselves are not independent since
  $\langle P(\Gamma^l | r ) P(\Gamma^r | r) \rangle
 \neq \langle P(\Gamma^l |r) \rangle \langle P( \Gamma^r |r) \rangle $.
The distant correlation of wavefunctions in the crossover 
regime \cite{FE96} follows from this relation.
The universal conductance peaks distributions $P_\lambda(g)$
 for  one-channel  symmetric leads are shown
in Fig. \ref{fig2} for several values of $\lambda$ in the crossover regime. 
 A good approximation 
 (see left inset in Fig. \ref{fig2}) is
$P_\lambda(g) \approx P(g |r_0)$, where 
 $r_0=r_0(\lambda)$ (see right inset in Fig.  \ref{fig2}) is determined 
by finding the best fit to the exact
$P_\lambda(\Gamma)$ for a single channel.  $r_0$ can also be 
estimated by  $\langle r \rangle$ 
 (solid line in the right  inset in Fig. \ref{fig2}).
 
To test our RMT predictions for the crossover conductance distributions, 
 we used the conformal billiard \cite{BR86} (threaded by an Aharonov-Bohm
flux line $\Phi$)
whose shape  is determined by the image of the unit circle
in the complex $z$-plane under the conformal mapping
$w(z) = (z + bz^2 + be^{i\delta}z^3)/\sqrt{1 + 5b^2}$. 
We collected statistics from several uncorrelated and fully chaotic billiards
with   $\delta=\pi/2$ and various values of $b$.
Semiclassical considerations for weak fields \cite{BG95} lead to a linear
 relation  $\lambda=\Phi/\Phi_{cr}$. 
 $\Phi_{cr}$ is the crossover flux given by
$\Phi_{cr}/\Phi_0 = (\pi/4) (\alpha_g^2 {\cal N})^{-1/4}$, 
where ${\cal N}$ is the number of electrons within the dot
and  $\alpha_g$ is a geometrical factor. 
Except for a constant factor, this crossover field has a similar
 expression to that of the correlation field \cite{AA96}.
Rather than calculating the geometrical factor semiclassically, 
 we  determined the exact relation
between $\lambda$ and $\Phi/\Phi_0$ by fitting 
 the Dyson-Mehta $\Delta_3$ spectral statistics to its
 known analytic form \cite{PM83,BG95}. 
The results for the lowest 300 eigenvalues are shown in the top
 inset in Fig. \ref{fig3}. 
For each lead we chose a sequence of $\Lambda$ equally spaced points
on the billiard boundary with spacing 
$ |\Delta {\bf r}|$.  To test our expression for the correlation matrix $M$,  
we have calculated  the averaged  wavefunction amplitude
correlations and found them to be in good agreement with 
$J_0(k |\Delta \mbox{\boldmath$r$}|)$ independent of the flux (see the 
 inset to the middle panel of Fig.  \ref{fig3}).
   Fig. \ref{fig3} compares the conductance distributions in the conformal
 billiard with  $\Phi/\Phi_0=0.04$  (histograms) to the theoretical predictions
(using $\lambda= 0.16$ from the inset) for three cases: one-channel leads,
 two-channel leads with $k |\Delta {\bf r}|=0.5$  
and four-channel leads with $k |\Delta {\bf r}|=2.4$. 
In each case we also show the limiting distributions for conserved
 and fully broken time-reversal symmetry. The billiard calculations
 are in good agreement with the intermediate distributions.

In conclusion, we have derived in closed form the width and conductance
peak distributions in a chaotic quantum dot  for leads with any number of
possibly correlated channels in the crossover regime
from orthogonal to unitary symmetry.
The distributions depend only on the symmetry-breaking transition
 parameter and the channel correlation matrix $M$ in each lead. 
This work was supported in part by the DOE, Grant No.
DE-FG02-91ER40608 and by the NSF, Grant No. PHY94-07194.
 NDW was supported by the European Human Capital
and Mobility Programme and by NSERC Canada.
YA acknowledges useful discussions with S. Tomsovic.

\begin{figure}


\caption
{ The distributions $P(\psi_{\mu R})$ (wider)  and $P(\psi_{\mu I})$ (narrower) 
of the real  and imaginary parts  
 of an eigenvector component in the principal frame for $\lambda=0.1$.
Histograms:  simulations of the RMT ensemble (\protect\ref{TranEns}); 
dashed lines: a Gaussian approximation; solid lines: the exact analytic results.
The  right inset shows  the components of
  a typical  vector  in the complex amplitude plane. 
The   left  inset is the r-distribution $P_\lambda(r)$ for $\lambda=0.1$, 
where the solid line is the analytic result (\protect\ref{rDist}) and the
 histogram is from RMT simulations.}
\label{fig1}

\vspace{3 mm}

\caption
{ Conductance peak distributions $P_\lambda(g)$ vs. $\log g$ in
 the crossover  from conserved to broken time-reversal symmetry for one-channel
symmetric leads:
$\lambda=0$ (GOE, dashed),
$0.1,0.25,0.5$ (solid lines) and $\lambda \gg1$ (GUE, dot-dashed).
Curve maxima increase with $\lambda$. 
  The  left  inset compares the Gaussian approximation for
 $P(g)$ (dashed)  with the exact result (solid)  for $\lambda=0.1$.
 Shown in the right inset  is
 $r_0(\lambda)$ (circles) and $\langle r \rangle_\lambda$ versus $\lambda$ 
(solid line).}
\label{fig2}

\vspace{3 mm}

\caption
{Conductance peak distributions $P(g)$ in the conformal billiard (histogram) 
for $\Lambda$ point-contact
symmetric leads and flux of $\Phi/\Phi_0=0.04$ for several values of $\Lambda $
and $k|\Delta {\bf r}|$.
 In each case we show the analytic predictions (solid lines) as well as
  the GOE (dashed) 
and  GUE (dot-dashed)  limits. The top inset  describes the transition
 parameter $\lambda$ as a function of magnetic flux $\Phi/\Phi_0$. 
 The inset in the middle panel shows the spatial correlations of the
 eigenfunctions for $\Phi/\Phi_0 =
0.02, 0.06$ and $0.10$  (diamonds, pluses and x's, respectively). 
The solid line is the theoretical prediction.}
\label{fig3}
\end{figure}

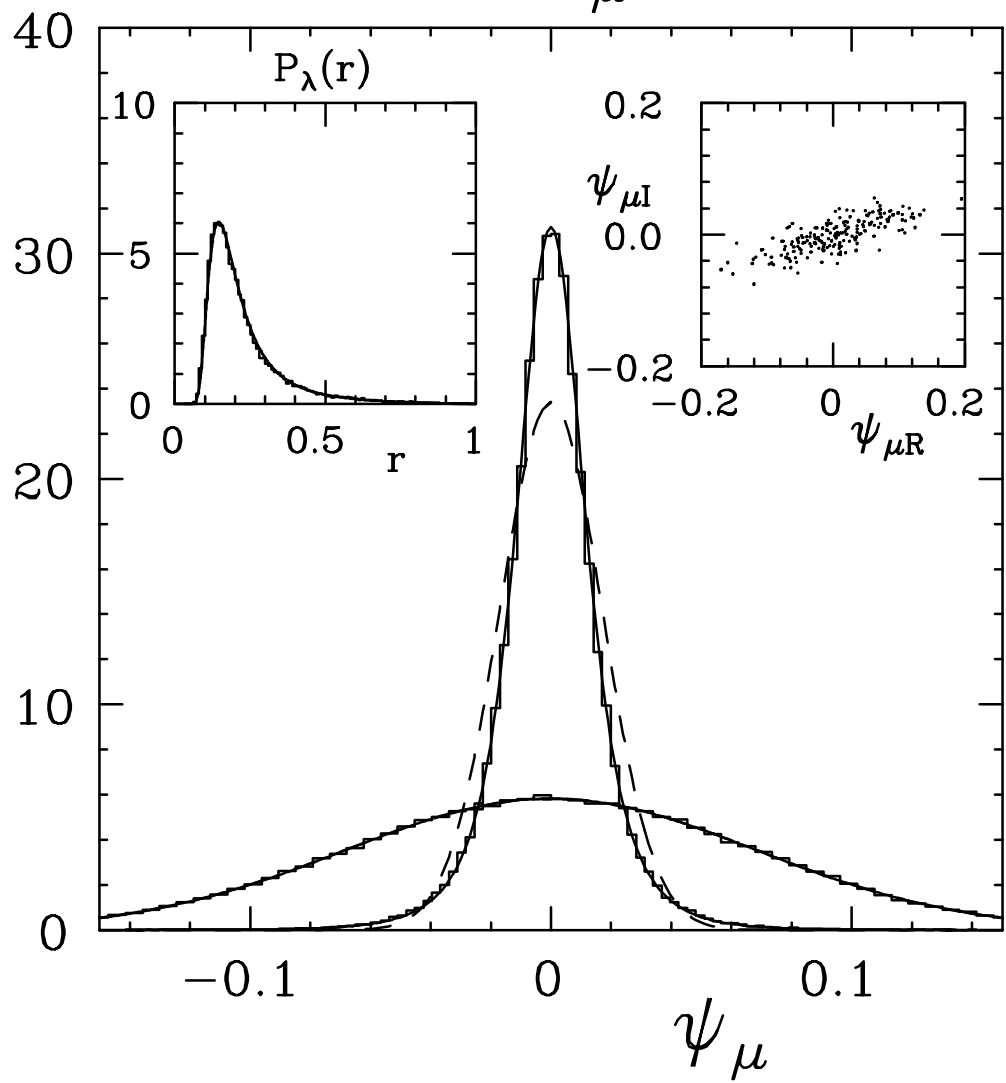
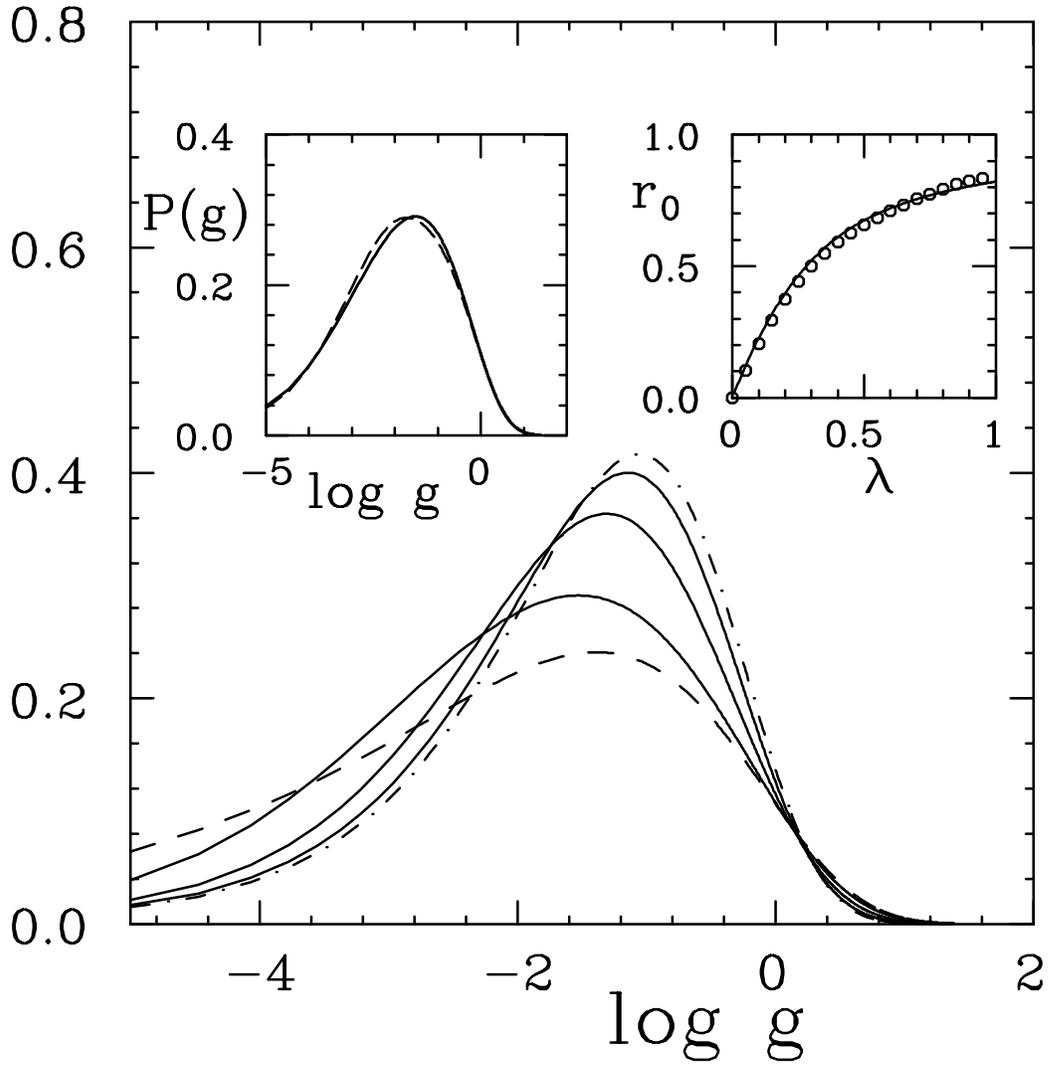
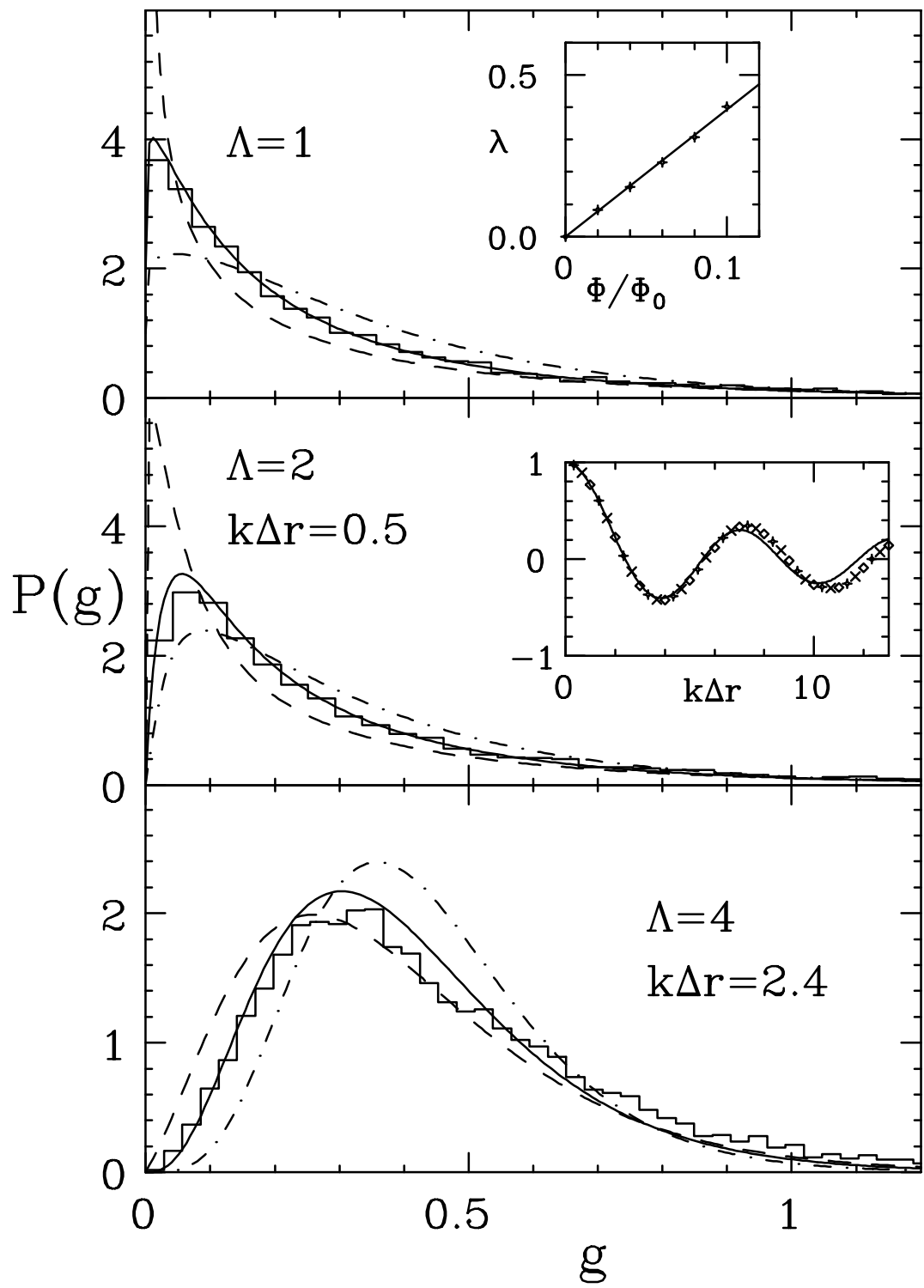
\end{document}